\documentstyle[twoside,fleqn,espcrc2]{article}

\newcommand{\AmS}{{\protect\the\textfont2
  A\kern-.1667em\lower.5ex\hbox{M}\kern-.125emS}}

\def\hat{\widehat}

\def\qbar{{\overline{q}}}
\def\slash#1{\mbox{$\not \!\! #1$}}
\newcommand{\Dslash}{{\slash D}}
\def\lrDslash{{\stackrel{\leftrightarrow}{\slash D}}}
\def\rDslash{{\overrightarrow{\slash D}}}
\def\lDslash{{\overleftarrow{\slash D}}}
\newcommand{\nfrac}[2]{{\textstyle\frac{#1}{#2}}}

\title{Non-perturbative improvement of 
operators with Wilson fermions\thanks{Talk presented by S. Sharpe.}}

\author{C.~Dawson\address{Dept. of Physics, Univ. of Southampton,
Southampton SO17 1BJ, UK}$\!$, 
G.~Martinelli\address{Dip. di Fisica, Univ. di Roma ``La Sapienza'',
P.le A. Moro 2, I-00185 Roma, Italy}$\!$,
G.C.~Rossi\address{Dip. di Fisica, Univ. di Roma ``Tor Vergata'',
Via della Ricerca Scientifica 1, I-00133 Roma, Italy}$\!$, 
C.T.~Sachrajda$^{\rm a}$,~ 
S.~Sharpe\address{Physics Dept., Univ. of Washington,
Seattle WA 98195, USA}$\!$,
M.~Talevi\address{Dept. of Physics \& Astronomy, 
Univ. of Edinburgh,
The King's Buildings, 
Edinburgh EH9 3JZ, UK}$\!$ 
and M. Testa$^{\rm b}$
}

\begin{document}

\begin{abstract}
We outline two methods of constructing improved composite operators using
Wilson fermions.
\end{abstract}


\maketitle

\section{Overview}

A major source of errors in present lattice calculations is
the use of a finite lattice spacing, $a$.
This is particularly true for the calculations of properties of
hadrons containing $c$ and $b$ quarks,
which contain discretization errors proportional to $a m_c$ and
$a m_b$, respectively.
Since lattice calculations have the potential to make significant
contributions to the study of D and B physics, it is
important to control such errors.

One approach to solving this problem is
the Symanzik improvement program.
This allows one to remove discretization errors order by order in $a$.
This program has been pursued by the ALPHA collaboration,
who have used chiral Ward Identities to determine,
non-perturbatively, the on-shell improved action,
the improved vector current including $a m$ terms,
and the improved axial current, scalar and pseudoscalar densities
in the chiral limit \cite{alphath,alphaex,luscher}.
Ward Identities for non-degenerate quarks allow the
determination of some of the $O(am)$ improvement terms,
at least in the quenched approximation \cite{divitiis}.
These methods do not, however, allow one to determine 
the improved pseudoscalar density, axial current or tensor density
away from the chiral limit.

To do so we have developed two methods which remove all errors of $O(a)$,
including those of $O(a m)$, from the matrix elements of bilinears.
These involve (1)
matching correlators of bilinears to their continuum form at
short Euclidean distances, and (2)
matching quark and gluon correlators to their continuum form at
large Euclidean momenta. 
Both rely on the restoration of chiral symmetry at short distances,
and both work for quenched and full QCD.
The first method requires only on-shell improvement,
i.e. improvement of physical quantities, 
while the second requires improvement of off-shell
correlators at an intermediate stage.\footnote{%
Throughout we use ``improved'' to denote quantities in which
all errors proportional to $a$ (multiplied by any power of $\ln a$)
have been removed. Errors of $O(a^2)$ remain.}
Details of the first method are given in Ref.~\cite{paper1}.
Details of the second method will be forthcoming, along with results
of a numerical pilot study \cite{paper2}. 
Here we provide a sketch focusing on some important issues.

\section{Improvement Program}

We need first to improve the action ~\cite{alphath}.
In continuum notation, the Wilson action,
\begin{equation}
S_{\rm W} = \int_x\left[ \nfrac{1}{2 g_0^2}\,
 {\rm Tr}(F_{\mu\nu} F_{\mu\nu}) 
+ \, \overline q (\Dslash + m_0) q \right] \,,
\end{equation}
has $O(a)$ errors due to the derivative.
%\begin{equation}
%\Dslash = 
%\nfrac{1}{2} \left(\gamma_\mu [\nabla^*_\mu+\nabla_\mu] 
%- a \nabla^*_\nu \nabla_\nu \right) \,.
%\end{equation}
Improvement requires the addition of all dimension 5 operators
allowed by the symmetries,
% of $S_{\rm W}$:
\begin{eqnarray}
\lefteqn{{\cal L}_{d=5} = - \nfrac{i}{4} {c_{SW}}\, a\, 
\overline q \sigma_{\mu\nu}F_{\mu\nu} q }\nonumber\\
&&\mbox{} + {b_g}
\, a\, m\, {\rm Tr}(F_{\mu\nu} F_{\mu\nu})/(2 g_0^2) 
- {b_m}\, a\, m^2\, \overline q q \nonumber\\
&&\mbox{} + {c'_1}\, a\, m\, \overline q (\lrDslash + m_0) q
+ {c'_2}\, a\, \overline q (\lrDslash + m_0)^2 q 
\,,
\end{eqnarray}
with appropriately chosen, $g_0$-dependent, 
coefficients \cite{alphath}.
Here $m=m_0-m_c(g_0)$ is proportional to the physical mass,
with $m_c$ the critical mass.
%Note that the gluon action is already improved at this order.

The r\^ole of the various terms is as follows:\\
\noindent
1. The ${\bf c_{SW}}$ term improves dimensionless physical quantities.
Its determination is clearly essential,
and has been carried out in Ref.~\cite{alphaex}.
\\
\noindent
2. The coefficients ${\bf b_g}$ and ${\bf b_m}$ remove
$O(a m)$ terms from the relation between 
bare and renormalized couplings and masses, respectively.
For example, $b_g$ introduces a mass-dependence in the effective gauge
coupling,
$g_{\rm eff}^2 = g_0^2 (1 - {b_g} am)$,
in such a way that the lattice spacing remains fixed as one varies $m$
at fixed $g_0$.
We will need to determine $b_g$ in our first method \cite{paper1}.
\\
\noindent
3. The coefficients ${\bf c'_1}$ and ${\bf c'_2}$ 
are only needed for off-shell improvement.
One way to see this is to note that
%, since they multiply terms vanishing by the equations of motion, 
they can be removed by a change of quark variables
(and an appropriate shift in $g_0$). 
This affects external sources, but not the spectrum.
In fact, one can ignore these terms also when doing off-shell
improvement, because they can be absorbed by a suitable change in
the improvement coefficients for quark fields \cite{paper2}.

The next step is to improve the operators themselves.
On-shell improvement has been discussed in Ref. \cite{alphath};
off-shell improvement requires additional terms.
For example, the improved form of the bare pseudoscalar density 
$P=\qbar\gamma_5 q$ is (for degenerate quarks)
\begin{eqnarray}
\lefteqn{\widehat P(x) = {Z_P(g_0^2,\mu a)} (1 + {b_P}am) \left( P(x) 
\vphantom{\left[\rDslash\right]}
\right.}\\
&& \left. + a {c'_P} 
%\underbrace{
\qbar \left[\gamma_5 (\rDslash\!+\!m_0)\!+\! (-\lDslash\!+\!m_0) 
\gamma_5\right] q(x)
%}_{{E_{P}}} 
\right).
\nonumber
\end{eqnarray}
In addition to the two on-shell improvement coefficients, $Z_P$ and $b_P$,
there is new coefficient, $c'_P$.
This multiplies an operator which vanishes by the equations of motion,
and so does not contribute to on-shell matrix elements.
%Thus we need to keep $c'_P$ only for our second method.
The pattern is the same for other bilinears---each has a single
additional off-shell term \cite{paper2}.

%The improved axial current is
%\begin{equation}
%\widehat A_\mu(x) = {Z_A(g_0^2)} (1 + {b_A}\,a\,m) \left(
%\underbrace{\qbar(x)\gamma_\mu\gamma_5 q(x)}_{{A_\mu(x)}}
%+ a {c_A} \partial_\mu P(x)
%+ a {c'_A} E_{A} \right)
%\,,
%\end{equation}
%where $E_A$ is of identical form to $E_P$, with $\gamma_5$ replaced
%with $\gamma_\mu\gamma_5$. Here there are three on-shell coefficients,
%and a single additional off-shell coefficient, $c'_A$.

\section{Gauge-Invariant Method}

This method involves only on-shell quantities, and so we do not
need consider the off-shell coefficients such as $c'_P$.
We discuss the example of the pseudoscalar density---details for
the other bilinears can be found in \cite{paper1}.

To determine $Z_P$ and $b_P$ we require that the Euclidean
two-point function of the improved lattice operator 
(evaluated using the on-shell improved action),
\begin{eqnarray}
\hat G_P(x) &=& \langle \hat P(x) \hat P(0) \rangle \\
&=& {Z_P}^2 (1 + {b_P} a m)^2 
\langle P(x) P(0) \rangle
\,, \label{eq:gpres}
\end{eqnarray}
agrees with the continuum result up to $O(a^2)$. 
The latter can determined, at short distances, using the OPE
\begin{eqnarray}
\lefteqn{G_P^{\rm cont}(x) =
{1 - 2 \alpha_s \gamma_P \ln (x \mu)+ \dots 
\over 2 \pi^4 x^6} }\\
&& \times
\big[1 + O(m^2 x^2) 
+ O(m \Lambda_{\rm QCD}^3 x^4,\Lambda_{\rm QCD}^4 x^4)
\big] 
\,. \nonumber 
\end{eqnarray}
The term on the first line is the coefficient function of the unit
operator, with $\mu$ the renormalization point
and $\gamma_P$ the one-loop anomalous dimension.
%It diverges as $1/x^6$, up to logarithms.
The correction terms on the second line are, respectively, the
perturbative contribution due to the quark mass,
and the non-perturbative terms due to the operators $\bar\psi\psi$
and $F^2$ respectively.
The crucial point is that chiral symmetry is restored at short distances 
(i.e. there is no $m$ dependence if power corrections can be ignored).

%Note that the off-shell improvement term proportional to $c'_P$ does
%not contribute as long as $|x| > 0$.

Equating $\hat G_P$ and $G_P^{\rm cont}$ in the chiral limit yields $Z_P^2$
in the chosen renormalization scheme.
Demanding that $\hat G_P$ contain no
terms linear in the quark mass determines $b_P$, i.e.
\begin{equation}
(1 + b_P a m)^2 = G_P(x;m=0)/G_P(x;m) 
\,, \label{eq:bpcond}
\end{equation}
where $G_P$ is the bare lattice two-point function.
We emphasize that the determination of $b_P$
is non-perturbative, and independent of that of $Z_P$.

There is one subtlety in the determination of $b_P$.
The condition (\ref{eq:bpcond})
requires that the physical distance $x$ not depend on $m$.
This in turn requires the determination of $b_g$, as discussed above.
Note that this is only an issue for full QCD---$b_g=0$ in the quenched
approximation.
To determine $b_g$ non-perturbatively one must hold fixed
a physical quantity which itself has no dependence on $m$.
The choice we suggest is the force between a heavy $q-\qbar$ pair
at short distances. See Ref.~\cite{paper1} for details.

The determination of $Z_P$ 
requires that there exist a window at short distances:
$a \ll |x| \ll \Lambda_{\rm QCD}^{-1}$.
This ensures the smallness of both higher order discretization errors 
of $O(a^2/x^2)$ and non-perturbative contributions.
In addition, to determine $b_P$, one must work in a region where 
$m^2 x^2 \ll ma$ and $m\Lambda_{\rm QCD}^3x^4 \ll ma$,
so that the dominant mass dependence is due to the $ma$ terms.
These conditions can be satisfied for sufficiently small $a$,
and for a suitable choice of $m$.
Based on experience with non-perturbative renormalization on quark states,
we expect that $a=0.05-0.1$ fm will be small enough to apply the method,
but numerical tests are clearly required to check this.

A potential practical problem is the 
need to have both short and long distances
on the same lattice. Short distances are needed for the determination
of the improvement coefficients, while long distances are required
in order to calculate matrix elements using the improved operators.
This problem is not particular to our method---to calculate 
renormalization constants such as $Z_P$ requires making contact with
perturbation theory and thus working at short distances.
In the Schr\"odinger functional method, this problem is overcome by
matching between lattices of different spacings.
We wish to stress that a similar approach is possible here.
In essence one calculates the improvement coefficients with a very small
lattice spacing, and then matches these coefficients onto the lattices with
larger spacing on which the matrix elements are calculated.
The details are sketched in \cite{paper1}.

\section{Gauge Non-invariant Method}

Our second idea is an extension to $O(a)$ of the non-perturbative
renormalization program of Ref.~\cite{npren}.
We require that quark and gluon correlators agree with their
renormalized continuum counterparts at large Euclidean momenta.
Two major complications arise in this extension.
First, since we are improving off-shell quantities, we need to include
the additional improvement coefficients such as $c'_P$.
Second, since we must fix the gauge, the improvement terms are no longer
constrained by the gauge symmetry, but rather by the lattice BRST symmetry.
This allows additional improvement terms which are
gauge non-invariant (or non-covariant). The only such term which appears
at $O(a)$, however, is a $\slash\partial q$ term in the improved quark field:
\[
\hat q = Z_q (1 + b_q a m) \left[ 1 + a c'_q (\Dslash+m_0) + 
a c_{NGI} \slash\partial \right] q
\,.
\]
%The $c_{NGI}$ term is manifestly gauge non-covariant.
Gauge invariant bilinears require non-invariant improvement terms only
at $O(a^2)$.

We can determine all the on-shell and off-shell improvement coefficients
by requiring that chirality violating form
factors vanish at large Euclidean momenta. 
For example, the amputated vertex of the improved pseudoscalar,
$\widehat P$, should, for large momenta, be
proportional to $\gamma_5$ with no $a (\slash p_1 - \slash p_2) \gamma_5$ term.
It turns out that the off-shell terms play an essential role in this
procedure: one must first determine the off-shell coefficients
in order to correctly determine all the on-shell coefficients.
A numerical test of this approach is underway. 
A similar method has
been suggested in Ref. \cite{rakow}, although without the inclusion
the gauge non-covariant $c_{NGI}$ term.

\section{Conclusions}

We have proposed two types of improvement condition, both non-perturbative, 
and sketched their use for bilinears. 
They provide an alternative to Ward Identities
for determining improvement coefficients in the chiral limit,
and they have the advantage of working also away from the chiral limit.
This is particularly important for applications involving heavy quarks.
The methods should be straightforward to generalize to more complicated
composite operators.
We are presently studying their numerical efficacy.

\end{document}